# Zero-shot counting with a dual-stream neural network model

Jessica A.F. Thompson[1,3,*], Hannah Sheahan[1], Tsvetomira Dumbalska[1], Julian Sandbrink[1], Manuela Piazza[2], Christopher Summerfield[1,**]

[1] Department of Experimental Psychology, University of Oxford, Oxford, UK
[2] University of Trento, Trento, Italy
[3] Lead contact
* Corresponding authors

Correspondence: jessica.thompson@psy.ox.ac.uk, christopher.summerfield@psy.ox.ac.uk

## Summary

Deep neural networks have provided a computational framework for understanding object recognition, grounded in the neurophysiology of the primate ventral stream, but fail to account for how we process relational aspects of a scene. For example, deep neural networks fail at problems that involve enumerating the number of elements in an array, a problem that in humans relies on parietal cortex. Here, we build a 'dual-stream' neural network model which, equipped with both dorsal and ventral streams, can generalise its counting ability to wholly novel items ('zero-shot' counting). In doing so, it forms spatial response fields and lognormal number codes that resemble those observed in macaque posterior parietal cortex. We use the dual-stream network to make successful predictions about behavioural studies of the human gaze during similar counting tasks.

Main text excluding methods: 5,525 words

## Introduction

The meaning of a visual scene depends on both its contents and its structure. The contents of a scene are the objects it contains. For example, in each panel of **Fig. 1a**, there are two salient objects: cats and bowls. The structure of a scene defines how the objects relate to each other. The meaning of each panel in **Fig. 1a** depends on whether there are more cats than bowls or vice versa, and whether the arrangement of objects is orderly or disorderly. The importance of object relations for understanding scene structure has been appreciated for at least a century, since the first investigations of Gestalt psychology [1].

Over recent years, we have learned a great deal about the computations that underlie the recognition of lone objects presented briefly at the fovea. Lesion and recording studies imply that object recognition relies on ventral visual regions of the primate brain [2,3]. The mapping of naturalistic images to semantic labels can be modelled as a feedforward cascade through successive processing layers of a neural network [4–8]. Deep convolutional networks trained with gradient descent to label images develop neural population codes that roughly match those observed in electrophysiology and neuroimaging studies of the primate ventral stream [9–12]. This success with modelling perception of scene *contents* notwithstanding, computational models of how the relational *structure* of a scene is processed remain much less mature [13–15].

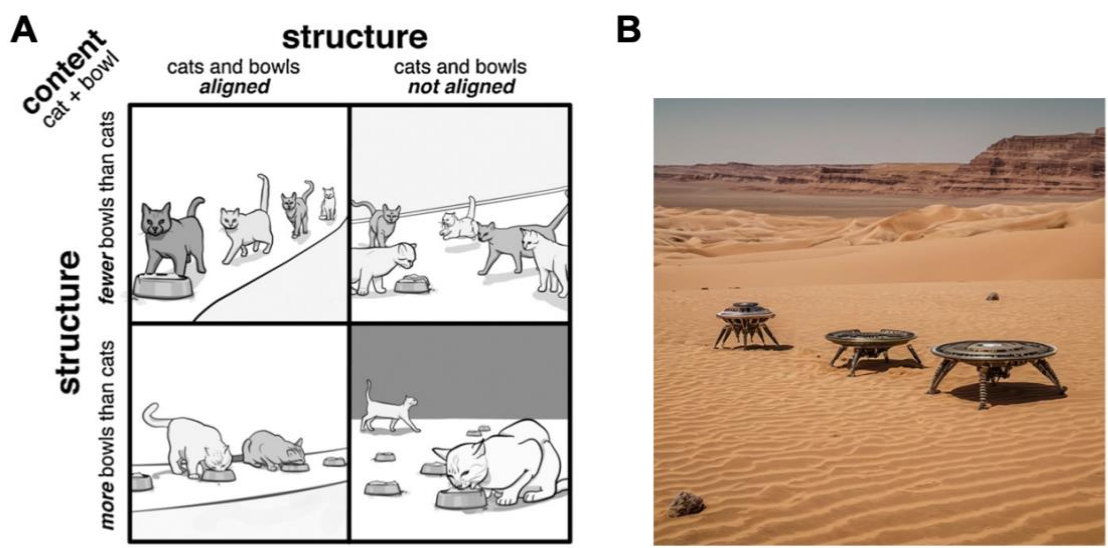

**Figure 1. A)** The meaning of a visual scene depends on both its content and structure. Image by Hannah Sheahan. **B)** Humans who master the cardinal principle will have no difficulty counting completely novel objects in novel contexts without needing any additional training examples—they generalise numerosity zero-shot. Image generated with https://www.img2go.com/.

One major challenge is that humans can immediately apprehend many relational aspects of a visual scene even if the objects it contains are wholly novel. For example, you may not recognise the objects in **Fig. 1b** but have no difficulty reporting that there are three of them (here, we call this phenomenon *zero-shot counting*). This ready ability to apprehend structure (object relations) without being able to recognise contents (object identity) is puzzling in the context of deep learning models, which often fail dramatically when probed about structural aspects of a novel scene. For example, neural networks struggle to identify whether two previously unseen objects are same or different [16]. In the case of counting, supervised learning of numerosity is severely disrupted when the objects being counted lie outside the training distribution [17,18]. Even very large generative models are prone to make structural errors in scene composition (including numerosity) when mapping text to images [19,20]. This implies that understanding scene structure relies on computational processes that not currently included in the canonical deep learning framework for modelling object recognition.

In humans, correctly inferring relations among objects in a scene depends on the integrity of dorsal stream structures, including the posterior parietal cortex (PPC). For example, patients with bilateral damage to posterior parietal cortex often have difficulty counting, comparing, or localising objects in a visual array. One possibility is that the biological brains evolved a visual system that factorises the processing of scene contents and structure into respective ventral and dorsal pathways [13]. In the current work, we describe a neural network model that implements this idea with a recurrent dual-streams architecture. We show that this network can solve the 'zero-shot counting' problem, and that in doing so, it develops neural representations that closely resemble those observed in the posterior parietal cortex of the nonhuman primate. The network also successfully predicts new behavioural results observed in human participants performing an eye tracking task that involves counting objects among distracters.

## Results

We call the problem we set out to solve zero-shot counting. It is operationalised as follows. The observer is asked to classify the number (1-5) of target items in a 2D grayscale image (**Fig. 2a**), potentially in the presence of up to two pre-specified distracter items. Both targets and distracters are alphanumeric characters embedded in a pixelated background. Foreground and background luminance values are sampled from Gaussian distributions whose means differ by at least 30% (**Fig. 2a**). To ensure that the task cannot be partially solved by counting the number of unique letters glimpsed, all target items within the array are the same letter. There is only one class of distracter (the letter A). During training, we sample targets from set $T_{train}$ (B,C,D,E). At test, with no further supervision signals provided, we evaluate the network counting performance for targets sampled from disjoint set $T_{test}$ (F,G,H,J). We also allow the distribution of mean luminance values to potentially vary between training and test (giving us $l_{train}$ and $l_{test}$).

### Zero-shot counting performance

In **Fig. 2b** we show the performance of a standard convolutional neural network (CNN) on this zero-shot counting task. We begin with the simplest case in which no distracters are present (*simple counting*). Whilst training on a set of images defined by $\{T_{train}, l_{train}\}$ we evaluate on new images drawn from the same distribution (validation) as well those drawn from a new distribution of luminance values (OOD luminance) new letters (OOD shape) or both (OOD both). The CNN successfully learns to count the items in the training data, and can generalise this to the validation set (accuracy = 99.9% +/- 0.01, chance = 25%), but not to the out-of-distribution (OOD) conditions $T_{test}$ (OOD luminance, accuracy = 81.6% +/- 17.7; OOD shape, accuracy = 72.1% +/- 6.7; OOD both, accuracy = 63.0% +/- 13.5). All mean accuracy and standard deviation values are calculated over 20 different random initialisations, and accuracy on all OOD test sets exceeded that for the validation set on each of these 20 seeds (One-sided Wilcoxon signed-rank tests, all $w$ = 210, $p$ < 0.001; all p-values Bonferroni corrected).

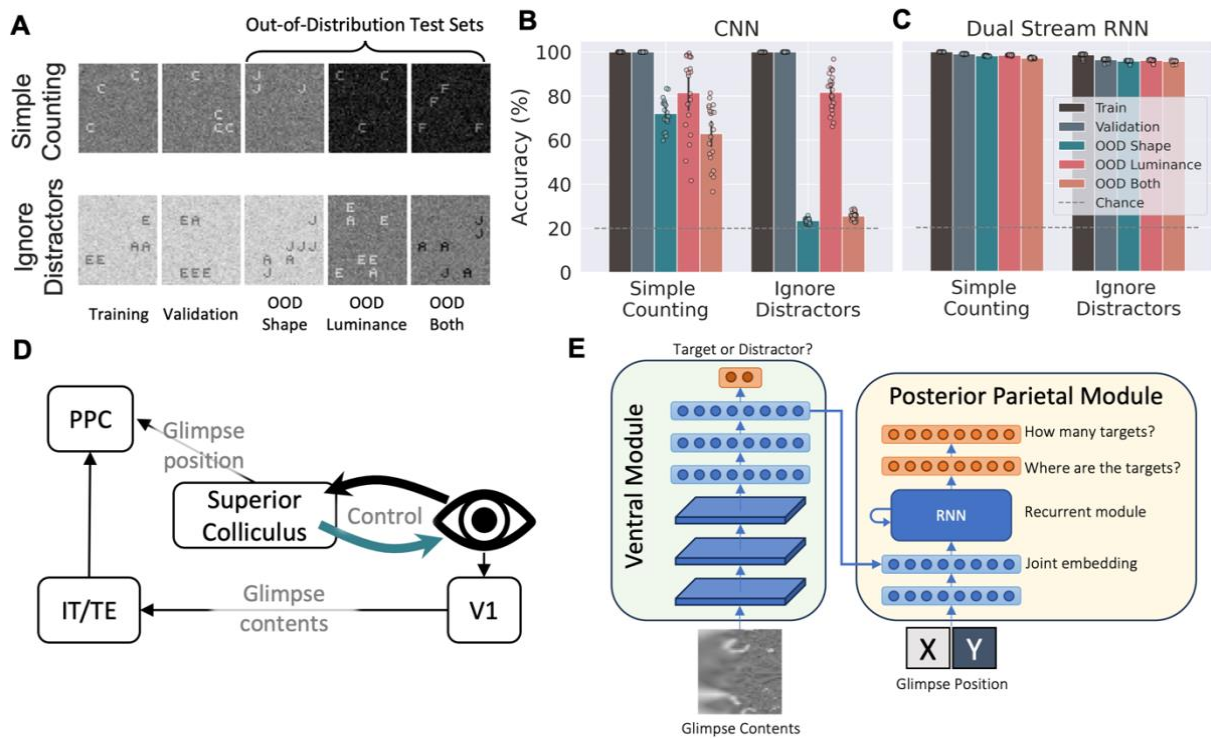

Figure 2. **A)** Example images for each task and dataset. For *simple counting*, there are no distracter items (no letters A). The validation set is independently sampled from the same distribution as the training set. The OOD test sets contain letters and/or luminances that were not present in the training set. **B)** Number classification accuracy for a convolutional neural network. Each dot represents one model run with a different random seed. Error bars indicate bootstrapped 95% confidence intervals. **C)** As B) but for the dual-stream RNN **D)** Schematic of the relevant components of the primate visual system. Efferent copies of motor instructions for intended eye movements propagate PPC via the superior colliculus. PPC integrates glimpse contents from the ventral stream and glimpse positions from superior colliculus. **E)** Dual-stream RNN architecture. A convolutional ventral module receives the foveated glimpse contents and is pretrained to distinguish target and distracter shapes. In parallel, a recurrent posterior parietal module receives the glimpse positions via a separate input stream. The recurrent module integrates the two streams over successive glimpses to produce a map of the spatial arrangement of target items in the array, from which the number of items is read out. Orange layers indicate where losses are calculated.

In the case where one or two distracters are present (*ignore distracters*), the CNN again performs well on the validation set (mean accuracy = 99.9% +/- 0.01), but its accuracy is dramatically reduced by OOD stimuli, especially in the OOD shape condition (23.5% +/- 0.10) and OOD both condition (25.5% +/- 1.8; accuracy in OOD luminance condition is 81.7% +/- 8.5). The CNN thus fails at zero-shot counting both with and without distracters. The finding that the CNN is perturbed by changes to irrelevant features of the image, such as luminance, might imply that CNNs solve the mapping problem by representing textural features [21,22]. However (without further constraints) CNNs do not naturally individuate objects as humans do when computing object relations, and thus struggle during counting of new objects.

The architecture of our proposed dual-stream recurrent neural network (RNN) model (shown in **Fig. 2e**) is inspired by the structure of the primate visual system, highlighted in **Fig. 2d**. Firstly, our network samples the image in a quasi-naturalistic way. Unlike the standard CNN, which receives the whole image at once as input, our network views each image through a sequence of spatially localised glimpses according to a biologically realistic gaze policy. It processes each glimpse with higher resolution at the locus of fixation, mimicking the primate fovea [23,24]. In the primate, glimpse contents are fed forward from thalamus and V1 to the ventral stream structures such as V4 and temporal cortex areas including inferior temporal cortex (IT/TE). Like others, we model this ventral stream processing with a convolutional architecture (in the *ignore distracters* condition, we pretrain this model to distinguish targets and distracters). The output of the ventral stream flows to higher association cortex, such as PPC which (along with PFC) is thought to integrate information across a sequence of glances in visual short-term memory [25]. To mimic this, in our model, outputs from the 'ventral stream' CNN are fed forward to a recurrent module that we equate with PPC. Recurrent computation allows information about number to be combined across glimpses (**Fig. 2e**, yellow box), and the proposed connectivity is consistent with known pathways from IT to PPC [26–28].

The key feature of our model, however, is that glimpse contents (what) is processed in parallel with glimpse location (where). We implement this dual-streams principle in the simplest way possible: on each saccade, we simply pass the $(x, y)$ location of the glimpse to the recurrent module. In the primate brain, we know that the superior colliculus (SC) encodes a topographic map of salient regions of visual space and computes a gaze vector, which is responsible for driving saccadic eye movements [29]. We also know that SC is reciprocally connected with PPC via the pulvinar [30–32], providing a putative pathway for the recurrent module to receive information about the current position of the eyes. Our dual-stream RNN includes two further layers that successively process outputs from the recurrent module. Its immediate readout layer (the penultimate layer of the network) is trained with an auxiliary loss to produce a spatial map of the location of target items in the image (we call this the 'map layer'). From the map layer, a linear read-out predicts the numerosity in the scene with a one-hot code (**Fig. 2e**).

Consistent with a previously described theory [13], we reasoned that this architecture would be able solve the zero-shot counting task because during training it would learn representations that explicitly combine information about glimpse contents ("what"; glimpse pixels from the ventral stream) and structure ("where"; glimpse location from the dorsal stream). We predicted that this would allow the network to generalise across scene structure (numerosity) even where scene contents (objects) were entirely novel [13]. In **Fig. 2c**, it can be seen that the dual-stream RNN is indeed able to solve the zero-shot counting problem. On the *simple counting* validation set, its performance is comparable to the CNN (99.0% +/- 0.1). However, unlike the CNN, it maintains this performance across OOD shapes (98.1% +/- 0.2), OOD luminances (98.4% +/- 0.2) and OOD both (97.1% +/- 0.3) conditions. It performs comparably in the *ignore distracters* condition (OOD shape 95.6% +/- 0.5; OOD luminance 96.0% +/- 0.6; OOD both 95.5% +/- 0.6). For the dual-stream RNN, the mean difference between validation performance and OOD both performance was less than 1 percentage point (0.8% +/- 0.3) for *simple counting* and similar for *ignore distracters* (1.9% +/- 0.3). By contrast, for the CNN, the mean difference between validation and OOD both performance was 37% +/-13.2 for *simple counting* and 74.5% +/- 1.7 for *ignore distracters*. Thus, the dual-stream RNN is able to solve the zero-shot counting task where the CCN is not.

## Ablations and controls

Next, we conducted control analyses that pinpoint those neural or computational features of our architecture that are critical for its success (**Fig. 3**). First, we carried out virtual lesion studies to examine what causal role dorsal and ventral stream played in network performance. We performed lesions by removing either glimpse contents (ventral stream lesions) or glimpse location (dorsal stream) inputs during both training and test. In the case of *simple counting*, dorsal stream lesions were more detrimental (reducing performance on OOD both to 48.0% +/- 2.5) than ventral stream lesions for accuracy (80.4% +/- 0.3 on OOD both). This is consistent with the finding from neuropsychological studies that PPC lesions lead to counting deficits, whereas temporal lobe lesions have a much milder impact [33–35]. By contrast, in the *ignore distracters* task, lesioning either the dorsal or ventral stream had a dramatic effect on performance. On the OOD both generalisation condition, classification accuracy was reduced to 42.1% +/- 0.4 by ventral stream lesions and to 52.7 +/- 0.3 by dorsal stream lesions (more than double the error rate observed in *simple counting*). Whilst we are not aware of neuropsychological data that directly supports this finding, there is good evidence that ventral stream lesions impair configural learning when object arrays become more complex [36,37].

Why does counting the number of objects in a scene (that is apprehended through a series of glimpses) require both what and where information? Intuitively, glimpse contents alone is often insufficient, especially if items in the array are identical. For example, if the network glimpses three items in succession, without auxiliary glimpse location inputs is unclear whether it has glimpsed three unique items ($i, j, k$) or glimpsed two before returning to the first item ($i, j, i$). However, gaze position alone is insufficient for counting, because saccades are not exclusively directed precisely to the centre of a single item, but often fall in an intermediate zone between two or more items, allowing the network to apprehend them in a single glimpse. Ventral stream lesions are especially detrimental in the *ignore distracters* task, because glimpse contents are needed to signal whether an item is target or distracter.

To study exactly how the network needs to combine what and where information in the task, we built a symbolic solver model which used a sequence of rules to classify the number of items given a set of glimpses (positions and contents) in a simplified version of the task. The solver worked by first attempting to infer the location of items from the glimpse positions alone, and then querying the glimpse contents only to resolve any remaining ambiguity. Thus, the number of queries to the glimpse contents provided an 'integration score' for each glimpsed image, indicating the degree to which the two input streams need to be integrated to solve the task. For example, if the set of glimpses contained only two unique positions in opposite corners of the image, it would be relatively clear from the glimpse positions alone that there are 2 items in the image. This would receive a low integration score. If, on the other hand, there are several glimpses clustered in a region that could reasonably contain 1, 2 or 3 items, the solver would need to inspect the contents of those glimpses to know for sure, prompting a higher integration score (for more details about the symbolic solver, see [38]). This solver allows us to pinpoint the impact of both ventral and dorsal lesions on the counting process (**Fig. S3**). When ablating the glimpse contents input, generalization performance scales inversely with integration score. The glimpsed images that the model struggles with in this one-stream setting are exactly those that the symbolic model identified as requiring an integration of both streams due to an ambiguity in the glimpse positions regarding item location.

Our dual-stream RNN differs architecturally from our control model (a vanilla CNN) in that it comes equipped with recurrent memory. Next, thus, we confirmed that recurrence alone was insufficient to solve the zero-shot counting problem. To test this, we created a one-stream version of the dual-stream RNN in which (like the CNN) the full image was input on each successive glimpse ("Whole image RNN" in **Fig. 3**). This control network was also significantly impaired on all three OOD conditions (*simple counting*: OOD luminance 54.6% +/- 4.1 ; OOD shape 48.4% +/- 1.7 ; OOD both 42.9% +/- 2.2). Removing the auxiliary map loss from the dual-stream RNN had a much more modest impact on performance (OOD both: *simple counting* 85.3% +/- 16.0, *ignore distracters* 93.9% +/- 0.2). Omitting the ventral stream objective, instead allowing all network parameters to be updated end-to-end with respect to the number and map objectives, greatly reduced its generalisation performance on the *ignore distracters* task (OOD both 42.9% +/- 11.3). Together, these analyses show that both the dorsal and the ventral stream are both necessary for solving the zero-shot counting task. We were unable to identify a trivial computational feature that can be added to a standard network to account for this success.

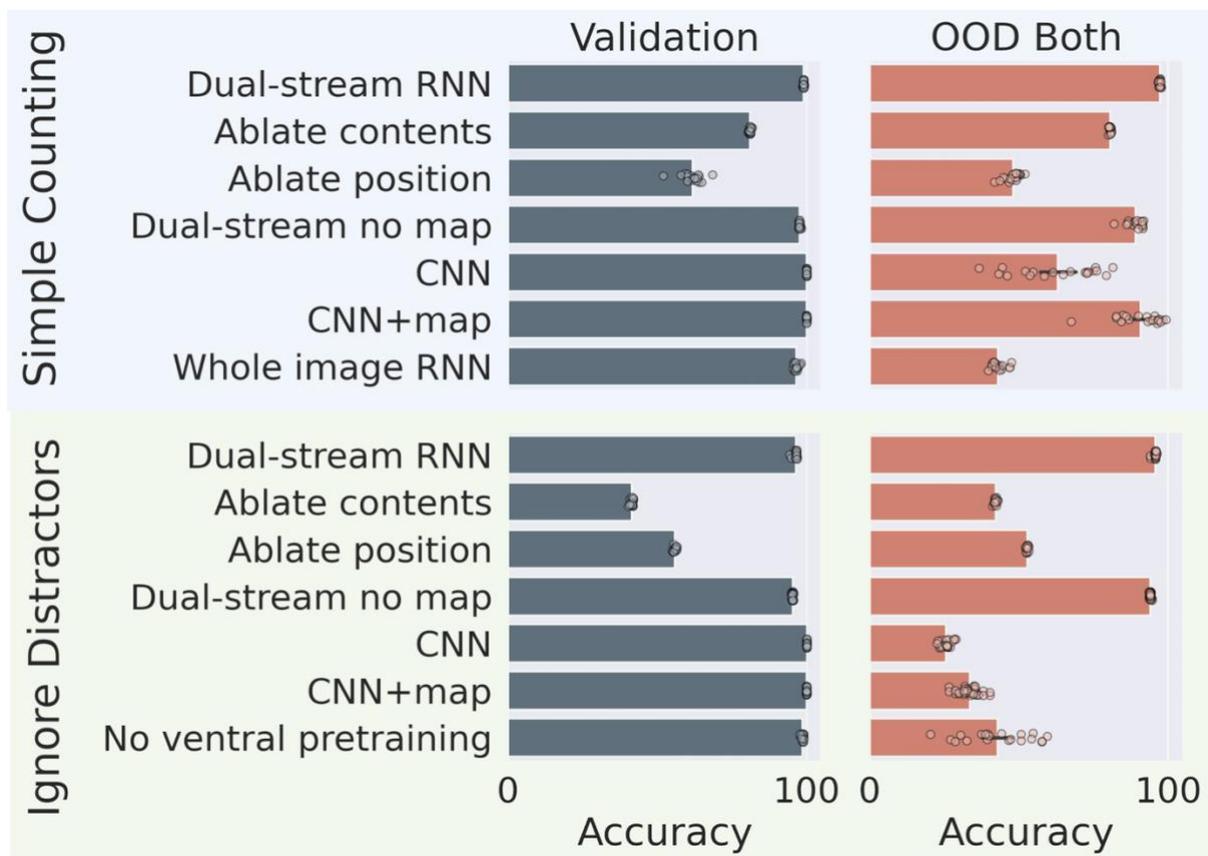

**Figure 3**. Accuracy on the validation set (left) and OOD both test set (right) for various control models and ablations. Each row corresponds to a particular training recipe or configuration of either the Dual-Stream RNN or the CNN baseline. Each dot is one model run with a different random seed. Error bars show bootstrapped 95% confidence intervals. When no error bar is present, the 95% CI was too small to be visible. Data are shown separately for the *simple counting* (upper panel) and *ignore distracters* (lower panel) conditions.

## Human behaviour

Next, we studied the behaviour of the dual-stream RNN across the trajectory of learning. As children learn to count objects in visual scenes, they often pass through discrete phases in which they are able to enumerate two, three or four items before grasping the general principle of cardinality. These phases are sequenced so that children may be "two knowers", "three knowers" and "four knowers" before graduating to become "cardinal principle knowers" [39]. We tested for this pattern in the behaviour of the dual-stream RNN over the course of training. In **Fig. 4a**, we show matrices which reveal the pattern of confusions (on the validation set) it makes for items with different numerosities for checkpoints lying at 25%, 50%, and 75% accuracy and at the end of the training run. In **Fig. 4b**, we plot training curves for arrays with a ground truth number of 1-5 items. It can be seen the dual-stream RNN, but not a CNN, learns to accurately classify arrays with a smaller number of items first.

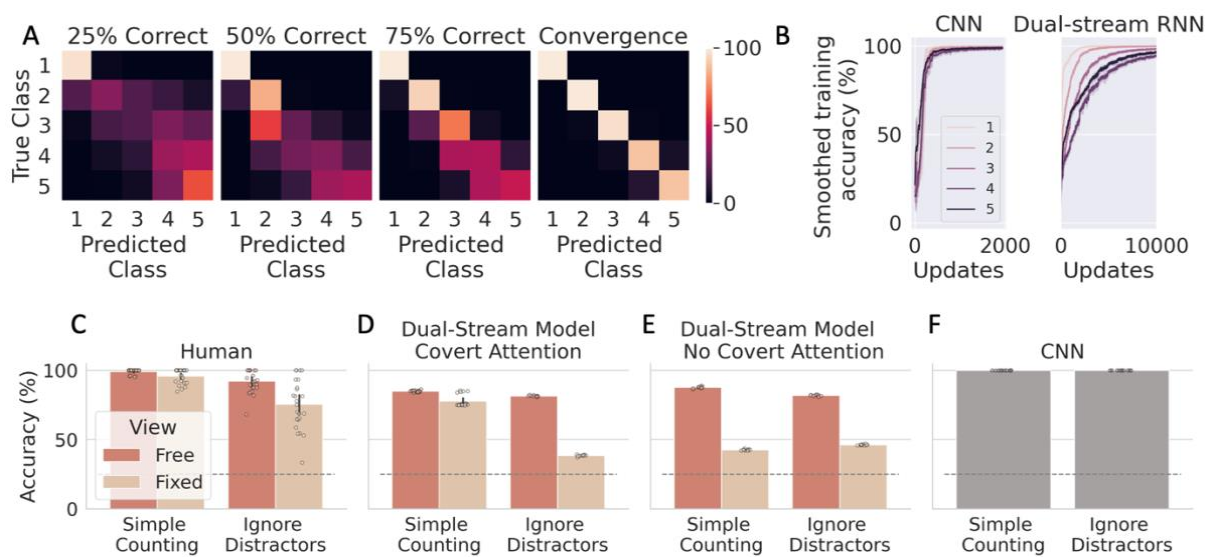

**Figure 4.** Comparisons with human behaviour. **A)** Confusion on the validation set (*ignore distracters*) for the dual-stream RNN at checkpoints throughout learning. Rows specify the actual number of items and columns specify the predicted number of items, according to the dual-stream RNN. The colour indicates the proportion of images of number class y that were predicted to be of number class x (light=high, dark=low). Rows sum to 100%. **B)** Smoothed mini-batch training accuracy per number class (1-5; for *simple counting*). This is the accuracy encountered before each weight update, which permits inspection of the learning curves at a fine temporal resolution. Error bars indicate standard deviation over five repetitions with different random seed. **C)** Human performance on the counting tasks. Each dot corresponds to a particular participants performance in one of four conditions. Dashed line indicates chance. Error bars are bootstrapped 95% confidence intervals. **D-E)** Validation accuracy of the dual-stream RNN with covert attention (D) and without (E). Each dot is one model run with a different random seed. **F)** Validation accuracy of the CNN. Here there is only one bar per task because there is no way of simulating 'Free' or 'Fixed' gaze conditions for the CNN. Each dot represents one model run from different random seed.

Our theory seemingly makes a counter-intuitive prediction: that humans' ability to count items in a visual array depends on our capacity to move our eyes. We know that eye movements and visual counting are linked. For example, patients with bilateral damage to PPC suffer from Balint's Syndrome, whose symptoms combine optic ataxia (disrupted saccadic eye movements) with simultanagnosia (an inability to perceive more than one object at a time) [40]. Nevertheless, people can perceive small numerosities in a single glance (such as when you read a number five off a die), an ability that is known as *subitising* [41]. This seems to present a challenge for our theory. We thus conducted an eye-tracking experiment involving human participants, to ask whether our network was able to predict patterns of human counting performance under free and fixed gaze.

We asked human participants (n = 24) to perform a visual counting task whilst we tracked their gaze position on the screen. We crossed task (*simple counting* vs. *ignore distracters*) with gaze (free vs. fixed) in a 2 × 2 within-subjects design. Stimuli contained 3-6 target items and, in the *ignore distractors* task, 1-3 distractor items. In free gaze blocks, participants could move their eyes as they wished, whereas in fixed gaze blocks, they were obliged to maintain their eyes within 100 pixels of central fixation during counting, or else the trial was aborted (and repeated at the end of the block). We found that in humans, fixing the gaze impaired performance to a much greater degree in the *ignore distracters* task than in *simple counting*. This observation was qualified by a two-way ANOVA on accuracy which revealed a significant interaction between task and gaze ($F(1, 92) = 10.23$, $p < 0.01$); there was a significant reduction in accuracy from free to fixed conditions on the *ignore distracters* task (mean diff = 16.6, Tukey HSD adjusted $p < 0.001$, family-wise error rate [FWER] = 0.05) but not the *simple counting* task (mean diff = 3.3, Tukey HSD adjusted $p = 0.6904$, FWER=0.05).

We simulated these data using our dual-stream RNN. On fixed trials, the network repeatedly glimpsed the centre of the screen, whereas free trials unfolded exactly as described above (for this simulation, we trained the network on an equal mixture of free- and fixed-viewing trials, so that neither gaze condition was out of distribution during test). We considered two settings: a *covert attention* setting where the network continues to receive putative gaze location information, even when it is forced to fixate centrally; and a *no covert attention* setting, where no gaze location information is offered on fixed trials. With covert attention, the network recreated the exact performance pattern observed in the human data – that performance on *ignore distracters* was affected by enforcing central fixation to a much greater degree than on *simple counting* (Mann-Whitney $U = 120$, $p < 0.001$). This was not the case if covert attention was removed (Mann-Whitney $U = 0$, $p = 0.99$). In this case, enforcing fixation affected both tasks almost equally. By contrast, the baseline CNN, which has no fovea, achieves 100% accuracy on the in-distribution validation set for both tasks and as such is unable to account for the pattern of errors that humans make in these conditions.

Neural coding

A natural next question is whether principles of neural coding in our network match those observed in the primate brain. We focussed on responses in the recurrent layer, which we equate with primate PPC. There are at least four canonical signatures for numerosity that are detectable in neural population codes recorded in macaque and human PPC (many of which are replicated in prefrontal regions) [42,43]. Firstly, neurons are tuned to number, and more neurons prefer the smallest and largest numbers in a discrimination, compared to those in between [44] (**Fig. 5a**). Secondly, when making numerosity judgments, some number selective units in intraparietal sulcus (IPS) display activity that ramps up over time, often with the steepest slopes for the largest numbers (**Fig. 5b**). Thirdly, neurons tend to code for number with approximately bell-shaped (Gaussian) tuning curves, and tuning width grows with number (log-normal number coding; Fig. 5c). Finally, at the population level, neural codes are low-dimensional, tracing out a neural "number line" which becomes visible when each number is expressed as a point in a space with just a few axes, derived with dimensionality reduction techniques (Fig. 5d).

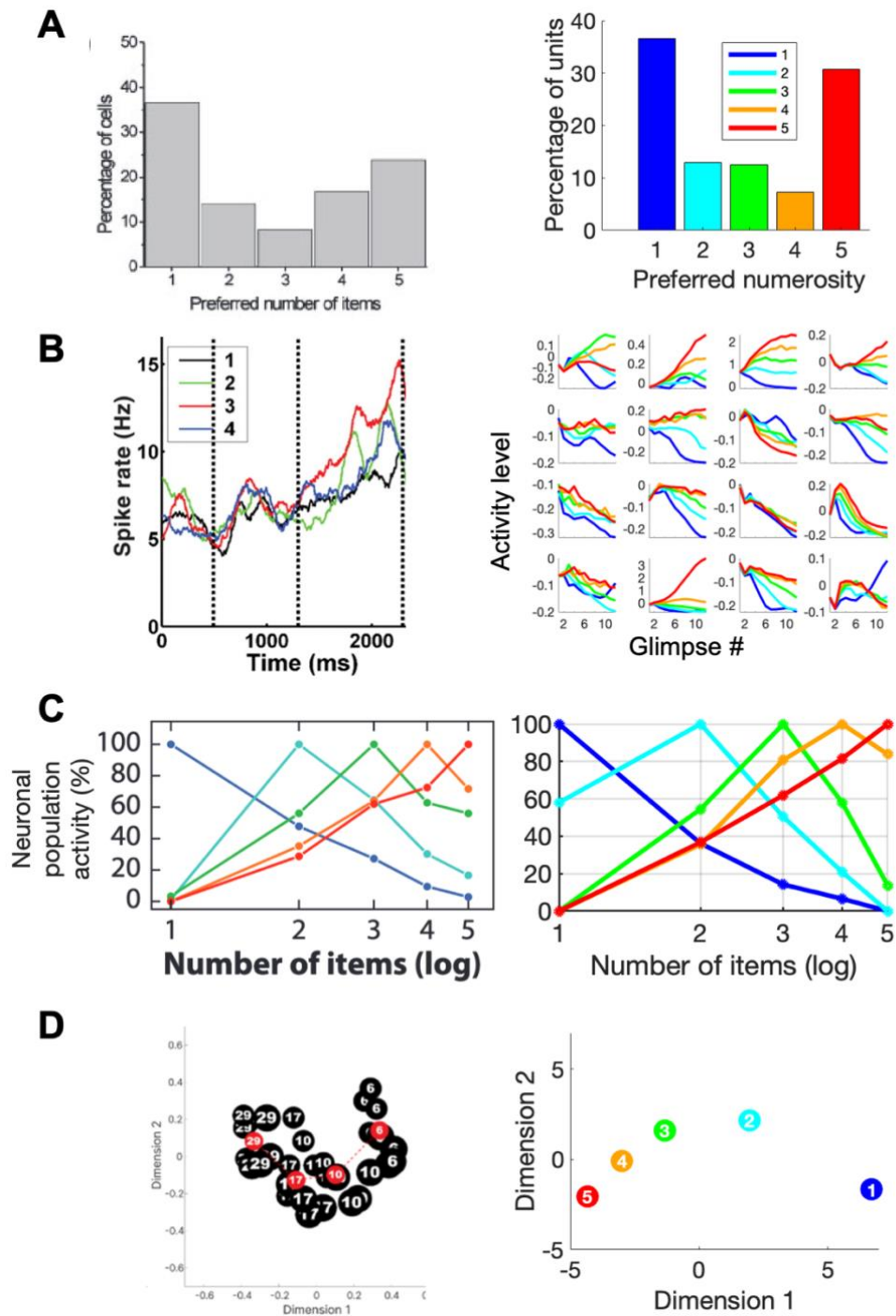

**Figure 5.** The dual-stream RNN mimics canonical signatures in neural population codes for number recorded in primate PPC. **A)** Left: Frequency distributions of the preferred numerosities for macaque PPC from [45]. Right: Frequency distributions of preferred numerosities among the recurrent units of the dual-stream RNN. **B)** Left: Spike density histogram for an IPS neuron from [46]. Right: Mean responses to each number class as a function of glimpse number for a random sample of units in the recurrent layer of the dual-stream RNN. **C)** Left: Gaussian tuning curves displaying lognormal number coding from [42]. Right: Average tuning of units in the recurrent layer of the dual-stream RNN. **D)** Left: MDS of BOLD activity from human IPS during dot counting from [47]. The black circles represent each stimulus labelled according to their numerosity (6, 10, 17, 29) and scaled in size to reflect the total area of the dots. The red circles indicate the average coordinates of each number. Right: MDS applied on the population activity of recurrent units in the dual-stream RNN.

We found that even without any hyperparameter tuning, the dual-stream RNN naturally recreates each of these neural coding motifs (**Fig. 5, right panels**). We show example tuning curves for numbers (1-5) in **Fig. S4**. Individual units develop preferences for different numbers, which tend to ramp up or down over successive saccades (**Fig. 5b**). This is similar to the pattern observed in macaque PPC when monkeys make numerosity judgments about arrays of dots [42,44,48]. When we plot the frequency distribution of preferred selectivity we can see that more cells prefer the extremes of the tested number range, a phenomenon which is also observed in PPC [45] (**Fig. 5a**). A salient feature of number coding in PPC is that when the tuning curves of cells are sorted and averaged, the coding is more precise for lower numbers, resulting in a characteristic "log-normal" code [42]. When we perform the same analysis, we see an identical phenomenon (**Fig. 5c**). Indeed, our average tuning curves were better fit by a model in which tuning curves were Gaussian in the space of $log(n)$ rather than $n$ itself (F-ratio test, F(1, 10) = 21.6, p < 0.001) (**Fig. S5**). Moreover, it is well known from macroscopic recordings that the neural similarity of the population response to number is well described by a single dimension, known as the "number line" [47,49]. We show the multidimensional scaling (MDS) projection of the population activity in 2D in **Fig. 5d**, which reveals a curved number line. To test the dimensionality of the population data, we split trials into two halves, and attempted to systematically reconstruct one half from dimensionality-reduced versions of the other. **Fig. S6** shows the mean variance explained on the held out half, as a function of the dimensionality of the half used for training. The best reconstruction was obtained when just three dimensions remained. Therefore, we infer that the dual-stream RNN has learned a low dimensional representation with a mental number line in the first dimension.

Next, we examined the spatial representations that formed in the network. Cells in PPC exhibit spatial response fields, in which firing rates are elevated in a temporal window surrounding a saccade made to that location [50,51]. In **Fig. 6a** we show examples of randomly selected spatial response fields for neurons in the recurrent layer. These closely resemble those observed in PPC, and also exhibit well-known properties of PPC response fields, such as the fact that neurons tuned to more eccentric locations have broader fields [52]. When we fit Gaussians to each observed spatial response field, we observed a positive correlation between the eccentricity of the best fitting mean (Euclidean distance to image centre) and the best fitting standard deviation (r = 0.20, p < 0.001) (**Fig. 6b**).

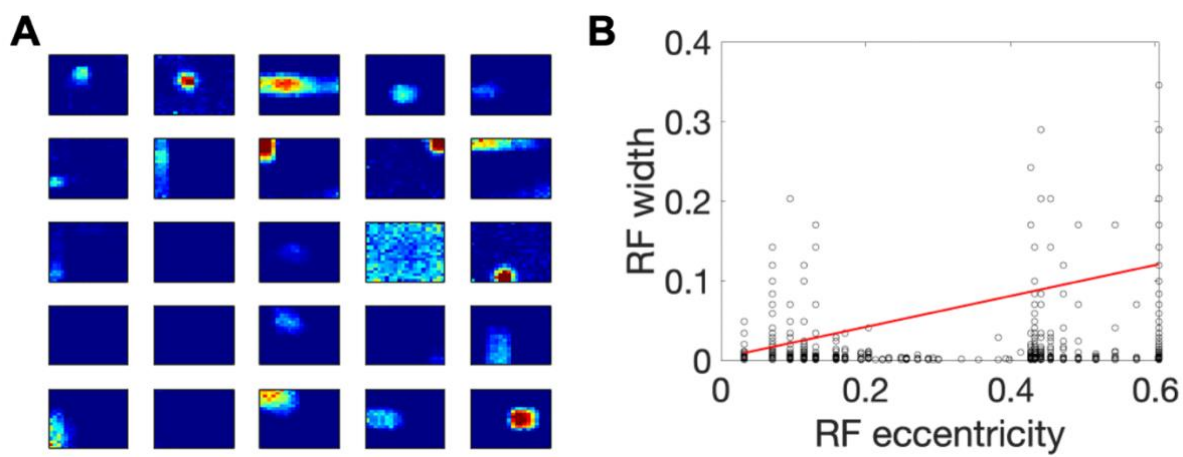

Figure 6. A) Example spatial response fields (RFs) of units in the recurrent layer of the dual-stream RNN. Warmer colours signal higher values. B) The width of the RFs of each unit in the recurrent layer of the dual-stream RNN as a function the eccentricity of the RF. Red line shows line of best fit.

## Discussion

The findings described here support an *enactive* view of cognition, in which motor signals (here, eye movements) are inputs to, as well as outputs from, neural computation. In our model, efferent copy is routed back as an input to the deep network, allowing it to learn representations that multiplex the structure and contents of a visual scene. This idea is similar to that proposed by emerging theories of learning in the hippocampal-entorhinal system, where the structure state spaces may be learned by taking actions and observing state transitions in an allocentric frame of reference [53]. Our model also helps unpack the puzzling relationship between space, number, and attention in the dorsal stream structures such as PPC. We propose that neural systems learn to allocate objects to spatial locations in a visual scene by using attention to orient across a scene in structured ways, which allows agents to multiplex information about what and where when learning new representations. By orienting attention, we can learn explicit representations of space that are not tied to a fixed set of familiar objects. Whilst our theory emphasises overt attention (saccades), covert attention (in which an internal spatial focus of processing is systematically oriented without a gaze movement) is likely to play a significant role in this process. Indeed, the assumption that PPC receives information about the covertly attended location was necessary to account for the data from our eye tracking experiment.

We studied a very limited aspect of numerical cognition, which is the ability to enumerate a small number of novel items in a visual scene. Our findings bear only tangentially on the other ways that numbers may be used, such as for example in mathematical calculations involving symbolic digits. Whilst there is evidence for parietal involvement in arithmetic , our model does not attempt to capture this ability, which (unlike visual enumeration) is unique to humans, and involves additional learning about the meaning of numerical symbols [54]. Nor is our model optimised to capture judgements about approximate number that can be made when an array contains tens or even hundreds of items [55,56]. As a serial model, our work does not explain the constant reaction times for very small numbers (1–3 items) in the absence of distractors, which may be better captured by pattern matching mechanisms in the ventral stream [57]. We note, however, that, consistent with our model, reaction times and number of saccades have been shown to increase with number of items for the range tested in our human experiment (3–6 items) [41], and a recent systematic review and meta-analysis has established that visual attention is integral in subitizing [58].

Many previous investigations of visual numerosity in deep neural networks have primarily been concerned with an innate or intuitive sense of visual number, which allows many animals and human infants to discriminate sets with different numerosity even without explicit training. In neural networks, this 'number sense' has been identified with the presence of 'number detector' units that are responsive to number while being relatively insensitive to other features like the shape, size, and spacing of the items. Such number detectors have been found in networks trained on visual object recognition [59,60], unsupervised objectives [61,62], or networks that are not trained at all (randomly initialised networks) [63,64]. This body of work has primarily been concerned with innate neural circuitry that is hypothesized to serve approximate numerical comparisons up to 20 or 30 items. In our work, on the other hand, we explicitly train our networks to perform exact enumeration of small numbers and we are primarily interested in the human ability to generalise systematically after learning to count—recognising learned relations (numerosity) in novel scenes. Innate number detectors are insufficient to explain this 'structure learning', which requires experience in humans and is not displayed by standard deep networks trained to classify exact numerosity [18,65–67].

The experiments and stimuli described here are deliberately made simple. Our goal is not to solve large-scale engineering challenges in computer vision, but to use deep networks as a vehicle for implementing a principle from neuroscience, and show how it can explain neural and behavioural phenomena observed in biological systems. Nevertheless, we believe that the principles described here could be scaled and may be useful for AI research. Indeed, glimpsing neural networks have previously been applied to tasks like visual object recognition [23,24,68]. One key outstanding question, which we leave for future work, is whether the approach described here could be used to help counting in naturalistic images (e.g. 3D tabletop scenes [69]). Another question is whether our approach extends to other Gestalt principles, and can be deployed to explain the human ability to judge relations of proximity, similarity, enclosure, symmetry and continuity with wholly novel objects. Finally, we focus on eye movements but the principle described here is more general, and could in theory extend to reaching movements, which no doubt help teach children about the structure of peri-personal space. Indeed, manual pointing in children seems to play an important role in learning [70,71] and has previously been employed in an RNN model of counting [72].

# Methods

Stimuli and task sets

We synthesised grayscale images (42 x 48 pixels) containing 1–5 target items (alphanumeric characters). For the *ignore distracters* task, an additional 0–2 distracter items were included. The primary objective was simply to report the number of target items in the image, either with (*ignore distracters*) or without (*simple counting*) potential distracter items. Characters lay on an invisible 6 x 6 grid, were of constant size (5 pixels tall, 4 pixels wide), and were never overlapping. N items were assigned to spatial locations in the image by randomly choosing N of the 36 possible grid locations. All target items within one image were the same character and had the same mean luminance value. Gaussian noise with standard deviation of 0.05 was added to both the background and foreground pixels.

For each task, we generated five datasets—one for training and four for testing. These datasets are summarized in Table 1. The five number classes were evenly represented in each dataset (and perfectly crossed with number of distracters in *ignore distracters*). The target shape, mean background luminance, and mean foreground luminance were sampled randomly from the set of target characters and set of mean luminances for that dataset. In the training set, target shapes were sampled from {B, C, D, E} and mean luminances from {0.1, 0.4, 0.7}. In any test set, these stimulus parameters were either the same as in the training or sampled from non-overlapping sets {F, G, H, J} and {0.3, 0.6, 0.9}. The distracter shape was always the character 'A'.

Table 1: Training and test set parameters. Dataset parameters are the same for both *simple counting* and *ignore distracters* except that no distracters are present in the *simple counting* images.

| Dataset | Target Shapes | Distracter Shape | Mean Luminances | # Images |
|---|---|---|---|---|
| Training | B, C, D, E | A | 0.1, 0.4, 0.7 | 100k |
| Validation | B, C, D, E | A | 0.1, 0.4, 0.7 | 5k |
| OOD Shape | F, G, H, J | A | 0.1, 0.4, 0.7 | 5k |
| OOD Luminance | B, C, D, E | A | 0.3, 0.6, 0.9 | 5k |
| OOD Both | F, G, H, J | A | 0.3, 0.6, 0.9 | 5k |

Glimpsing

Like the primate visual system, our dual-stream RNN apprehends an image via a sequence of foveated glimpses. A sequence of fixation points is generated according to a fixed, saliency-based saccadic policy. The saliency map is composed from a mixture of Gaussians with one Gaussian centred on each item. Our saccadic policy samples fixation points from the saliency map, subject to the constraint that each Gaussian is sampled at least once (i.e., each item is glimpsed at least once). To tune our saccadic policy, we used human eye tracking data from an independent study which presented similar images containing alphanumeric characters. First, we set our number of fixations to 12 based on the observation that, during a 3 second viewing period, human participants made 12 saccades on average. Second, we matched the standard deviation (in both x and y directions) of the fixation coordinates in the reference frame of the nearest item. In the human data, this standard deviation was 5% of the total image width/height. We adjusted the dispersion of the Gaussians in the saliency map to replicate this property in the simulated fixations (**Fig. S2**).

The mapping from the retina to the cortex in humans is well described as a log-polar transformation in which the horizontal and vertical axes in the retina are transformed into polar axes in the cortex: angle and eccentricity (distance from fovea—log scaled) [73]. To prepare our glimpse contents, we simulated the retinal-to-cortical transformation as a log-polar transform centred on the fixation coordinate using the *warp_polar* function from the scikit-image python package (version 0.19.2). The radius of the circle that bounded the transformed area was $r = \sqrt{width^2 + height^2}$ such that the entire image would be included in a glimpse directed at the centre of the image. Points outside the boundaries of the image were filled according to the 'edge' interpolation mode which pads missing values with neighbouring values in the image. The output shape of the log-polar transform was set to be the same size as the original image. The log scaling of the eccentricity coordinates results in foveated glimpse contents in which the foveal region is magnified relative to the periphery.

Model architecture

Our dual-stream RNN, inspired by the parallel pathways of the primate visual system, receives both the glimpse positions (the fixation coordinates) and the glimpse contents (the log-polar transformed image) as separate inputs streams. Separate feed-forward layers produce equal-sized embeddings (512 units) of both glimpse positions and contents. These two embeddings are concatenated before being passed through another layer to produce a joint embedding (1024 units). For the *ignore distracters* task, the contents embedding layer is preceded by a ventral module consisting of three convolutional layers and two feedforward layers. The activations of the penultimate ventral layer are passed on to the contents embedding layer. The joint embedding layer is followed by a recurrent module. The recurrent module consists of three transformations (input to hidden, hidden to hidden, and hidden to out) which preserve the size of the representation (1024). The output of the recurrent module is passed to a feed forward layer with 36 units. We call this layer the "map layer" because we train it to reflect the spatial arrangement of target items in the image. A final linear readout layer (5 units) predicts the number of target items in the image. For the *simple counting* task, this results in 5,268,696 total trainable parameters. All activations functions are leaky rectified linear (slope=0.1) except for on the map layer where a sigmoid activation is used to get values between 0 and 1. The model architecture is depicted in **Fig. 2c** and detailed model parameters are listed in Table 1 and Table 2. This design embodies the hypothesis that efferent copies of signals pertaining to a viewer's orientation to a scene, e.g., eye movements, rather than merely outputs of a visual system are also inputs that support learning useful representations of space and number.

Our convolutional neural network baseline model consists of three convolutional layers and two fully connected layers before the number readout. All convolutional layers consist of 56 feature maps, use a stride of 1, and use no padding. The first convolutional layer uses a 3x3 kernel and subsequent convolutional layers use a 2x2 kernel. The first fully connected layer has 256 units and the last layers have 36 and 5 units respectively. This results in 5,454,461 total trainable parameters. Detailed network parameters are listed in Table 3.

Table 1: Trainable model parameters, dual-stream model for *simple counting*

| Layer | Shape | #Parameters |
|---:|:---:|:---:|
| contents embedding | [512, 2016] | 1,032,192 |
| location embedding | [512, 2] | 1,536 |
| joint embedding | [1024, 1024] | 1,048,576 |
| rnn.i2h | [1024, 2048] | 2,098,176 |
| rnn.h2o | [1024, 1024] | 1,049,600 |
| map | [36, 1024] | 36,900 |
| number readout | [5, 36] | 180 |
| Total params with biases | | **5,268,696** |

Table 2: Trainable model parameters, dual-stream model for *ignore distracters*

| Layer | Shape | #Parameters |
|---:|:---:|:---:|
| ventral conv1 | [10, 1, 3, 3] | 100 |
| ventral conv2 | [10, 10, 3, 3] | 910 |
| ventral.conv3 | [10, 10, 3, 3] | 910 |
| ventral.fc1 | [50, 15120] | 756050 |
| ventral.fc2 | [10, 50] | 510 |
| proximity readout | [2, 10] | 22 |
| contents embedding | [512, 10] | 5,632 |
| location embedding | [512, 2] | 1,536 |
| joint embedding | [1024, 1024] | 1,048,576 |
| rnn.i2h | [1024, 2048] | 2,098,176 |
| rnn.h2o | [1024, 1024] | 1,049,600 |
| map | [36, 1024] | 36,900 |
| number readout | [5, 36] | 180 |
| Total params with biases | | **4,999,102** |

Table 3: CNN trainable model parameters

| Layer | Shape | #Parameters |
|---:|:---:|:---:|
| conv1 | [56, 1, 3, 3] | 560 |
| conv2 | [56, 56, 2, 2] | 12,600 |
| conv3 | [56, 56, 2, 2] | 12,600 |
| fc1 | [256, 21168] | 5,419,264 |
| fc2/map | [36, 256] | 9,252 |
| number readout | [5, 36] | 185 |

| | |
|---|---|
| *Total params with biases* | 5,454,461 |

Model training
All networks were built and trained using the PyTorch python library [74] version 1.12.1 on a single NVIDIA Titan X Pascal GPU. Additional python packages *numpy*, *matplotlib*, *pandas*, *seaborn*, and *scipy* were used for data analysis and visualisation.

For the *ignore distracters* task, the total set of trainable model parameters of the dual-stream RNN $\theta$ can be divided into those that make up the ventral module $\mathcal{V}(c_g)$ where $c_g$ are the glimpse contents of a particular glimpse $g$ and those that make up the parietal module $\mathcal{P}(\mathcal{V}_{-1}(c), p)$, where $\mathcal{V}_{-1}$ indicates the output of the penultimate layer of the ventral module and $c$ and $p$ are the sequence of 12 glimpse contents and positions respectively.

$$\theta = \{\theta_{ventral}, \theta_{parietal}\}$$

These parameters are optimised with respect to three different objective functions: a shape recognition loss $\mathcal{L}_{shape}(\theta_{ventral})$, a spatial map loss $\mathcal{L}_{map}(\theta_{parietal})$, and a number classification loss $\mathcal{L}_{number}(\theta_{parietal})$. During a pretraining phase, the ventral module is trained to predict the proximity of a glimpse to any nearby target or distracter items.

$$\mathcal{L}_{shape}(\theta_{ventral}) = \text{MSE}\left(s_g, \mathcal{V}(c_g)\right)$$

The target vector $s_g$ for this shape recognition task is constructed as follows. For a particular glimpse $g$, we calculate the Euclidean distances $d_i$ from the fixation point to every item $i$ in the image within $3\sigma$ of the fixation point (where the dispersion of the isotropic Gaussians that make up the saliency map is $\sigma^2$). For each item within range, we calculate the proximity as $1 - d_i/3\sigma$. The final target vector $s$ consists of two values which correspond to the sum of the proximities for distracters and targets respectively. For example, a glimpse directed exactly at the centre of a target item with no other items in the vicinity would produce a proximity vector $s_g$ of [0, 1] indicating that this is a very 'targety' glimpse. A similar vector would also obtain for a glimpse directed in between two neighbouring target items with no other items in the vicinity. If instead the glimpse was directed between a target and a distracter item, the proximity vector would be approximately [0.5, 0.5]. The ventral module is pretrained on an independent dataset containing all letters and all luminances. The parameters of the ventral module are held fixed during subsequent training of the parietal module.

Recall that the penultimate layer of the dual-stream RNN has 36 units corresponding to the 36 image "slots"—the spatial locations spanned by the 6x6 grid where items may appear in the image. This map layer is supervised with a binary cross entropy loss to produce a binary map of where the target items appear in the image,

$$\mathcal{L}_{map}(\theta_{parietal}) = \text{BCE}(\mathbf{m}, \quad \mathcal{P}_{-1}(\mathcal{V}_{-1}(c), p))$$

where the map target vector $\boldsymbol{m}$ contains a 1 for every slot that contains a target item and a 0 otherwise and $\mathcal{P}_{-1}(\mathcal{V}_{-1}(\boldsymbol{c}),\boldsymbol{p})$ indicates the output of the penultimate layer of the parietal module (the map layer).

From this map representation, a final readout layer produces the numerosity prediction, on which a standard cross entropy classification loss is computed,

$$\mathcal{L}_{\text{number}}(\theta_{\text{parietal}}) = \text{CE}(\mathbf{n}, \quad \mathcal{P}(\mathcal{V}_{-1}(\boldsymbol{c}),\boldsymbol{p}))$$

where $\mathbf{n}$ is the number of target items in the image. The optimised objective function is simply the sum of the number loss and the auxiliary map loss.

$$\mathcal{L}(\theta_{parietal}) = \mathcal{L}_{number}(\theta_{parietal}) + \mathcal{L}_{map}(\theta_{parietal})$$

All models were optimised with the AdamW optimiser (Adaptive Moment Estimation with Decoupled Weight Decay Regularization) [75] with weight decay of 1e-5. For glimpsing models, the order of glimpses was randomised anew at the beginning of each epoch to effectively augment the dataset. For recurrent models, we clipped the gradient norm at 2 to stabilise learning. All models were trained with a batch size of 512 for 300 epochs with a starting learning rate of 0.001 and a scheduler that decays the learning rate by a factor of 0.7 every 15 epochs. All models were trained 20 times from different random initializations. Reported results are averaged over these repetitions.

Control models and ablations

*Ablate contents/position*
When ablating one input stream or the other, the input is simply omitted both at training and test time. In this setting, what is labeled the "joint embedding" in **Fig. 2e** is a function of either the glimpse contents *or* the glimpse positions, not both. These are thus one-stream models.

*Dual-stream no map*
To interrogate the role of the spatial map representation in the penultimate layer of the dual-stream RNN, in this condition, we omit the map term $\mathcal{L}_{map}(\theta_{parietal})$, updating the weights only with respect to the number objective $\mathcal{L}_{number}(\theta_{parietal})$. Whenever the map loss is not optimised, the sigmoid nonlinearity at the map layer is replaced with a leaky rectified linear (slope=0.1) to avoid vanishing gradients.

*CNN+map*
In this version of the CNN baseline, we add the map loss term to the training objective. As in the dual-stream RNN, we calculate the map loss on the penultimate layer of the CNN (the second fully connected layer).

*Whole image RNN*
This is a one-stream, non-glimpsing control model whose architecture is identical to the Ablate position model. Instead of receiving a sequence of foveated glimpse contents, it receives the whole image repeatedly. This tests the role of recurrence alone.

*No ventral pretraining*
Normally the parameters of the ventral module are pretrained on the shape recognition objective $\mathcal{L}_{shape}$ and then held fixed during the training of the parietal parameters with respect to the number of map objectives. In this control model, we do not pretrain the ventral module and we do not update with respect to the shape objective at all. Instead, all parameters $\theta$ are updated with respect to the number and map objectives during the main training phase.

$$\mathcal{L}(\theta) = \mathcal{L}_{number}(\theta) + \mathcal{L}_{map}(\theta)$$

Neural Coding Analyses

We analysed the responses of the 1024 units in the recurrent layer of the dual-stream RNN to the 5000 images in the OOD both test set (*simple counting*) and on each of the 12 glimpses per image. All analyses were performed in MATLAB (2023a, The MathWorks Inc., Natick, MA) using custom code.

Number Coding
For each unit and on each glimpse, we calculated the mean response to each number class to see how the single-unit response to numerosity evolved over glimpses. We calculated the preferred numerosity for each unit as the number class that elicited the largest mean response over all glimpses. Units were then grouped according to their preferred numerosity and number tuning curves were calculated as the mean response over glimpses within units that preferred the same numerosity. Gaussian curves were fit to these tuning profiles using MATLAB's *fit* function in the space of linear numerosity and log numerosity. To inspect the geometry of the population response, we first calculated the pairwise Euclidean distances between the mean response to each number class (averaged over glimpses) and then applied classical multidimensional scaling (MDS) on the resulting dissimilarity matrices. This produced the 2D representation displayed in **Fig. 5h**. To test the dimensionality of the population activity, we split the trials into two halves which were dimensionality reduced separately according to the same procedure described above except that the pairwise distances were calculated on the responses to each image rather than the mean responses to each number class. Then a multiple linear regression was trained to predict the first half of the dataset from its dimensionality reduced version. This trained model was then tested on the second half of the dataset. The dimensionality of the reduction that produces the highest variance explained provides an estimate of the dimensionality of the population activity. We repeated this process on 1000 different random splits of the dataset.

Spatial Coding
To calculate spatial response fields for each unit, we defined a tiling of the whole image minus a 5% border on all sides resulting in a 21x21 grid of 441 tiles. For each tile, we identified all glimpses whose position was within the bounds of the tile. Then, the spatial response field was calculated for each unit as the mean response to glimpses directed at each tile. We then used the *mvnpdf* MATLAB function to estimate a 2D normal probability function for each response field.

Human Experiment

Ethics Statement

This study received ethical approval from the Central University Research Ethics Committee of the University of Oxford. All participants provided informed written consent.

*Participants*

Twenty-six participants (age 24.96 ± 2.69; 42% female, 58% male) took part. All of them were students or researchers at the University of Oxford. They received £15 as compensation for their time (60-90 minutes). Two participants were excluded from the subsequent analyses due to data corruption.

*Eye tracking Environment and Setup*

Each participant completed the study in one session that lasted between one and one-and-a-half hours. The experimenter remained in the room and recalibrated the system once after the practice trials, and then again after every other block of trials, with recalibrations taking place approximately every 10 minutes. Participants were seated in a dark room approximately 60cm away from a computer monitor (60 Hz refresh rate, 1280*1024 resolution, 17' LCD). Participants rested their head on an adjustable chin and head rest. Their eye gaze position was monitored using SR Research EyeLink 1000 and recorded at 1000 Hz. Fixation events were detected automatically by the SR Research Software.

*Stimuli*

Stimuli were synthetic grey-scale images containing between three and six *targets* (either the characters C, E, F, J, K, S, U, or Z), and – in *ignore distracters* trials - between one and three *distracters* (A's). Background and foreground (letter) luminances were always 0.3, 0.6, or 0.9, chosen randomly apart from the constraint that the difference in luminance was fixed at 0.3. Otherwise, image generation parameters were the same as described above under *Task sets*. We generated independent stimuli sets for each participant and each condition. These stimuli sets consisted of 72 images in which all numerosities were represented equally (and crossed with number of distracters where applicable). We also generated separate stimuli sets for the training blocks – one per condition per participant.

*Task Procedure*

Visual stimuli were presented with Psychtoolbox-3 [76] for MATLAB. Each stimuli set was split over two experimental blocks of 36 trials. Each experimental session consisted of four training blocks of eight trials, followed by eight experimental blocks of 36 trials for a total of 288 experimental trials per participant. The core task for each trial was to count either all letters (*simple counting*) or all letters but A's (*ignore distracters*) present in an image. Each trial was subject to one of two viewing conditions. Participants were either allowed to move their eyes freely (free gaze), or they were asked to fixate on a red cross in the middle of the screen (fixed gaze). Fixed gaze trials were only accepted if the participant maintained central fixation (within a radius of 100 pixels of the centre of the screen) for the whole duration of the stimulus presentation. Rejected trials were appended at the end of the respective block, to ensure the complete datasets for each participant. The two tasks and viewing conditions were combined according to a 2x2 design, so there were four conditions overall.

Participants received instructions before the start of each block, indicating whether they can move their eyes and whether to count all letters or ignore A's. They confirmed they had understood by pressing any key, and then the block would begin. The four practice blocks were assigned to the four conditions in the following order: (1) *simple counting, free gaze*, (2) *ignore distracters, free gaze*, (3) *simple counting, fixed gaze*, (4) *ignore distracters, free gaze*. For the experimental blocks, condition order was randomly assigned, subject to the constraint that each condition appeared once in the first group of four blocks, and once in the second group. Thus, every participant completed one practice- and two experimental blocks for each condition.

Each trial followed the same structure: (1) A circle converged toward a fixation point in the screen's centre within two seconds. (2) Once the point disappeared, the stimulus was presented in a 728x728 square in the centre of the screen. It remained there until either two seconds elapsed, or the participant pressed spacebar to confirm that they were ready to respond. (3) Following presentation of a *mask image* to suppress iconic memory for 0.1 seconds and a gap of 0.5 seconds, (4) the *response screen* was displayed, which asked participants to press keyboard keys 3, 4, 5, or 6 to indicate the number of target items. Their response was displayed on the screen and they were asked to confirm their response by pressing the button again. Once the participant confirmed, they either received feedback (*correct* or *incorrect*) or, if they had moved their gaze too far away from the centre of the screen in a *fixed gaze* trial, a message reporting trial rejection for 1.5 seconds. After a brief interval of 0.5 seconds, the next trial followed.

*Behavioural Analysis*
Behavioural analyses were performed in Python version 3.10.8 (Python Software Foundation) using the statsmodels python package version 0.14.1. For each participant, the accuracy per condition was calculated as the number of correct trials out of the total number of trials in that condition. Performance was compared across conditions using a 2x2 ANOVA (Type III) and subsequent Tukey HSD Test (n=24). For five of the participants, 1–8 (out of 288) trials were missing due to a recording error. These trials did not belong to a particular condition, and thus none of the participants' data were excluded from the analysis.

## Code Availability
All code used to generate stimuli, train models, collect human behaviour, and analyse results is available at https://github.com/summerfieldlab/saccades.

## Acknowledgements
This work was supported by generous funding from the European Research Council (ERC Consolidator award 725937) and Special Grant Agreement No. 945539 (Human Brain Project SGA). Thanks to David McCaffrey and Adam Harris for early discussions.

## Author Contributions
Conceptualization, C.S., M.P., J.A.F.T., and H.S.; Methodology, J.A.F.T., T.D., and H.S.; Investigation, J.S., J.A.F.T., and T.D.; Writing – Original Draft, C.S., J.A.F.T., and J.S. ; Writing – Review & Editing, J.A.F.T., C.S., T.D., J.S., and M.P.; Funding Acquisition, C.S.; Resources, C.S.; Supervision, C.S. and J.A.F.T.

Declaration of interests

C.S. is employed by the UK Artificial Intelligence Safety Institute. H.S is employed by Google Deepmind.

Supplemental information

Document S1. Figures S1-S6

# Supplemental Figures

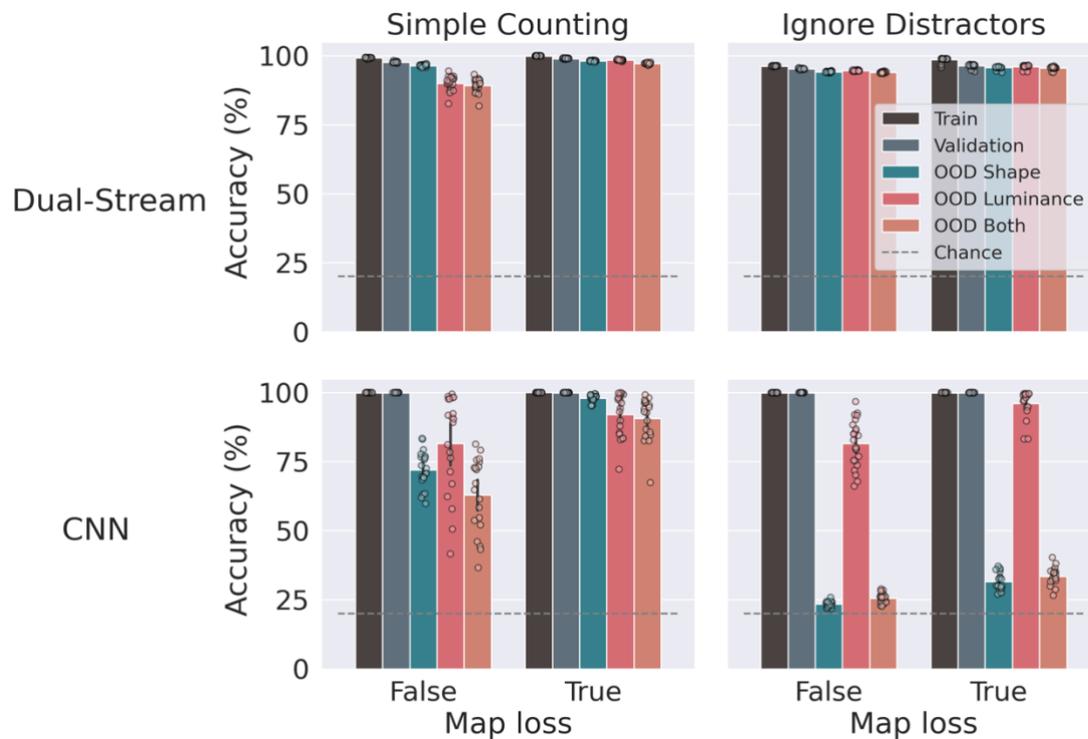

**Figure S1.** Effect of the auxiliary map objective. The training, validation, and out-of-distribution test accuracy when training with or without the auxiliary map loss term for the dual-stream model (top) and the baseline CNN (bottom) on the simple counting (left) and ignore distractors (right) tasks. Each dot corresponds to one of 20 models trained from different random seeds. Error bars show the bootstrapped 95% confidence interval around the mean accuracy over those 20 repetitions.

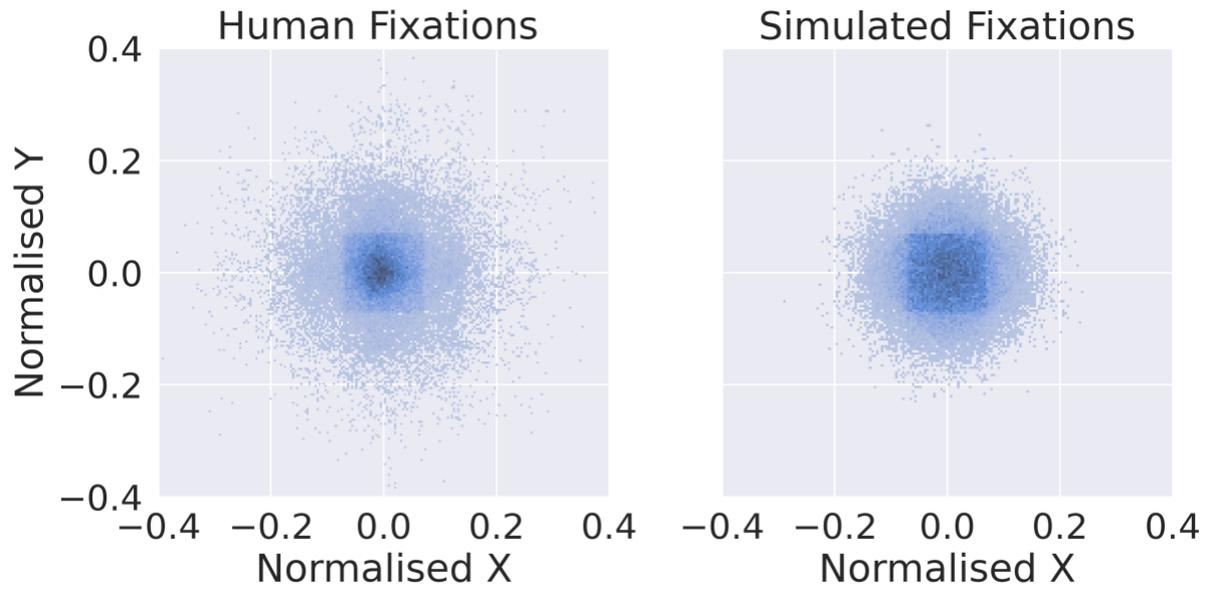

**Figure S2.** Histograms of fixation points in the reference frame of the nearest item for both human (left) and simulated (right) fixations. Axes are normalised by the size of the image. Darker colours indicate higher density. The saliency map from which the simulated fixations are sampled was configured so that the covariance of these two distributions was equal. In both, the standard deviation in x and in y is 5% of the total image.

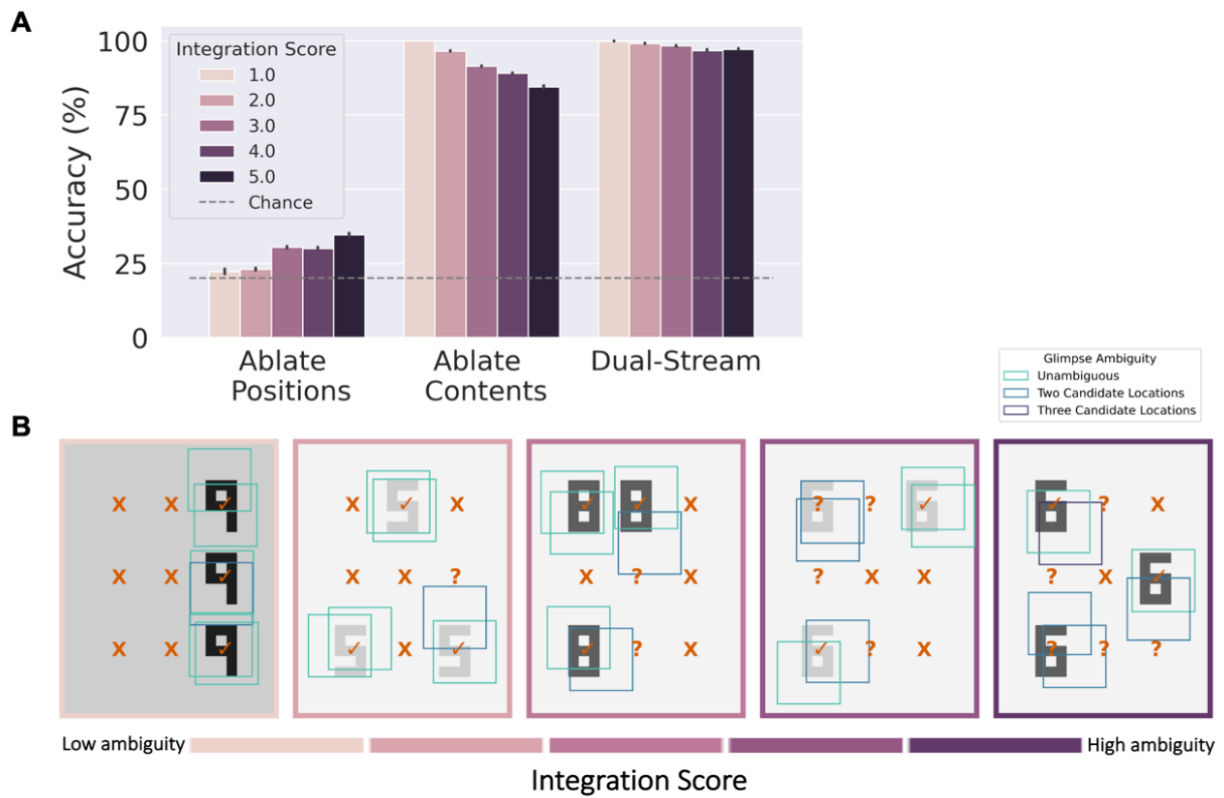

**Figure S3.** Integration score analysis to understand how the model combines both input streams. A) OOD both test performance for the dual-stream RNN when it receives only the gaze contents (Ablate Positions), only the gaze locations (Ablate Contents), or both input streams (Dual-Stream) divided by integration score. Integration score is calculated per glimpse sequence with a manual solver that attempts to infer the numerosity from a symbolic version of the glimpse contents and positions. The integration score is a function of how many times the solver needed to query the glimpse contents to infer the correct numerosity in a simplified version of the simple counting task. The integration score constitutes a prediction of which glimpse sequences will be most challenging to a uni-stream model without glimpse contents. Consistent with this prediction, we observed that the errors in the Ablate Contents condition (that are corrected in the Dual-Stream condition) are predominantly made on glimpse sequences with high integration score. B) Illustration of the symbolic manual solver. Each panel shows an example image with glimpses (blue boxes, coloured to denote ambiguity about item- location assignments) overlaid on items. Panels are arranged from left to right in increasing order of integration score (1-5). The orange ✓, X, and ? symbols respectively indicate spatial locations where the solver determined that there is, is not, or could be an item based on the glimpse positions alone. For an integration score of 1 (far left), there are no interrogatives because there is no ambiguity about where the items lie. See Thompson, Sheahan & Summerfield (2022) for more details about the symbolic solver.

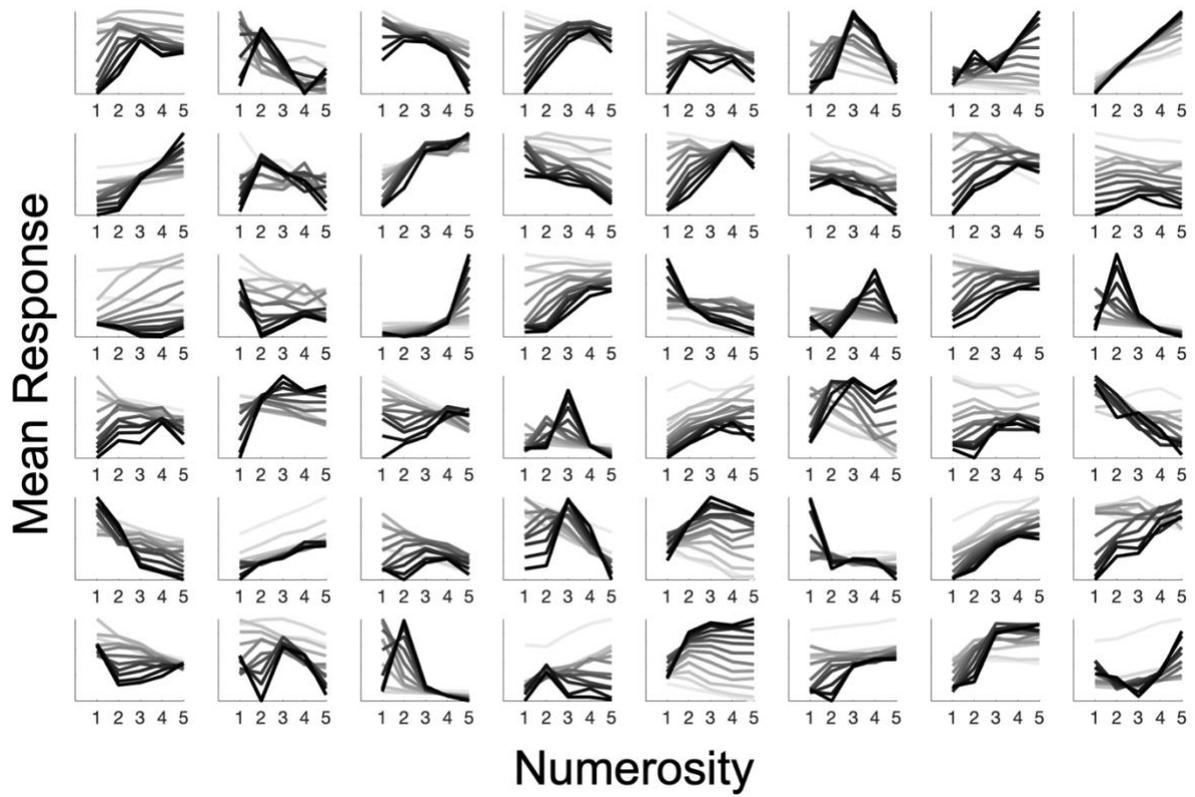

**Figure S4.** Tuning curves for individual units. Each subpanel corresponds to a different randomly selected unit in the recurrent layer of the dual-stream RNN trained on the simple counting task. Lines show the mean response to images of each number class in the OOD both test set. Darker lines correspond to later glimpses. Tuning curves generally become steeper and sharper over successive glimpses.

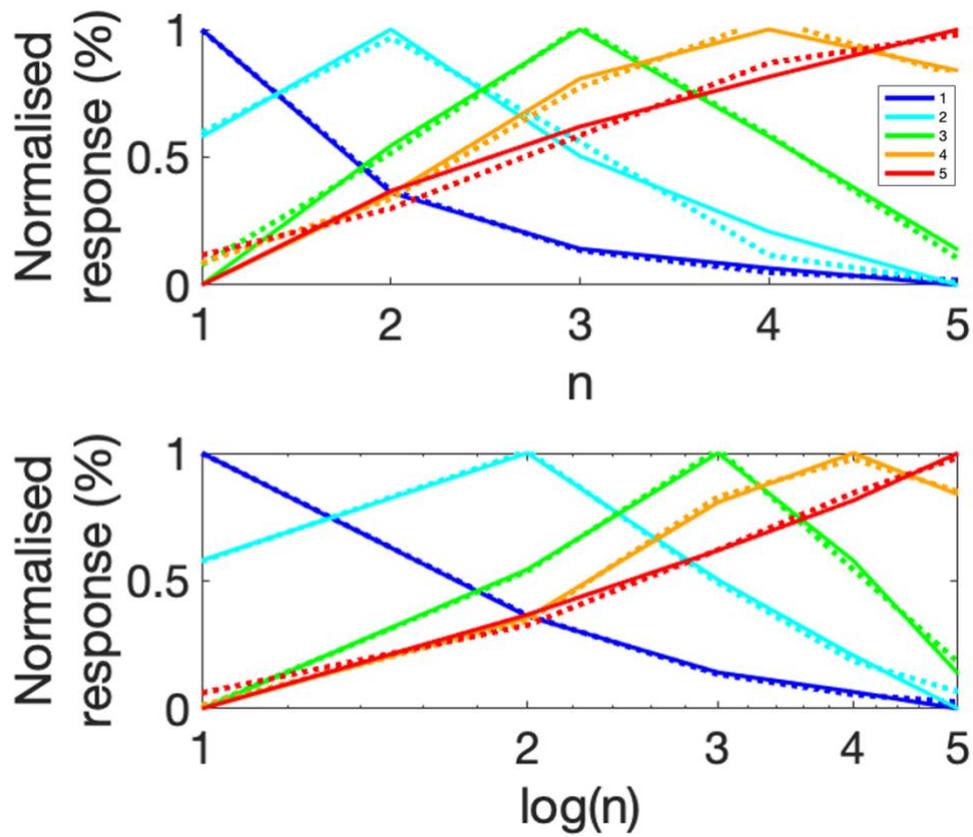

**Figure S5.** Average tuning curves. Colour indicates preferred numerosity. Unbroken lines show the mean response to images containing n items, averaged within units that prefer the same number and over successive glimpses. Dashed lines show the Gaussian fit, either in the space of n or log(n).

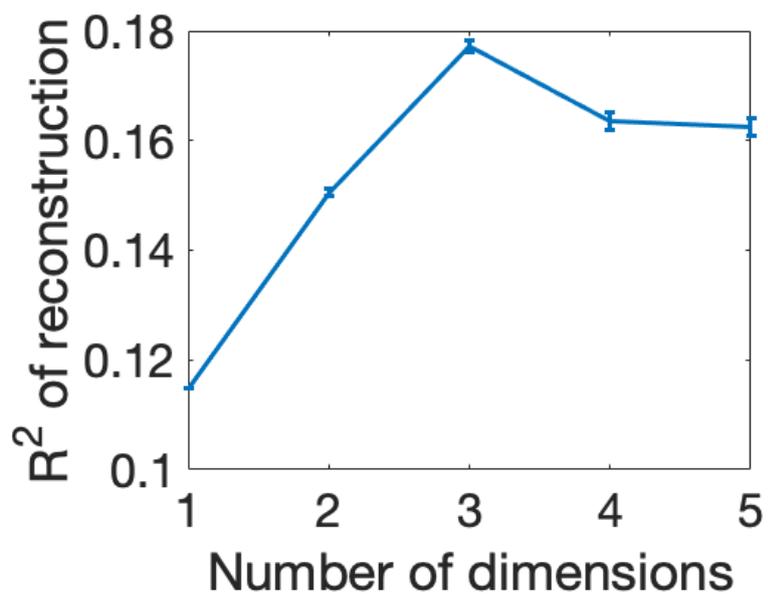

**Figure S6.** Analysis of the dimensionality of the representation in the recurrent layer of the dual-stream RNN. Variance explained when reconstructing the mean response to one half of the images with a linear model trained on a dimensionality reduced version of the other half, for a range of dimensionalities. R-squared values are averages over 1000 random split-halves and error bars indicate the standard error. Variance explained peaks when the dimensionality is reduced to three, suggesting that the subspace of the recurrent activations that is consistent over repeated splits of the data can be well captured by only three dimensions.